# Excitonic order in quantum materials: fingerprints, platforms and opportunities

Yande Que[1*], Clara Rebanal[2*], Liam Watson[3], Michael Fuhrer[4], Michał Papaj[5], Bent Weber**[1], and Iolanda Di Bernardo**[2,4,6]

[1]School of Physical and Mathematical Sciences, Nanyang Technological University, Singapore, 637371 Singapore
[2]Departamento de Física de la Materia Condensada, Universidad Autónoma de Madrid, Madrid, 28049 Spain
[3]Australian Synchrotron, Clayton 3168, Victoria, Australia
[4]School of Physics and Astronomy, Monash University, Clayton, 3800, VIC, Australia
[5]Department of Physics and Texas Center for Superconductivity (TcSUH) at the University of Houston, Houston, Texas 77204, USA
[6]Instituto de Física de la Materia Condensada, Madrid, 28049 Spain
[*]These authors contributed equally to this work
[**]e-mail: b.weber@ntu.edu.sg, iolanda.dibernardo@uam.es

## ABSTRACT

The exciton insulator (EI) is a unique many-body ground state of condensed, spontaneously formed excitons (electron-hole pairs) in equilibrium, distinct from conventional band or Mott insulators. Originally proposed over half a century ago, the concept has recently gained renewed experimental traction thanks to advances in spectroscopic resolution, ultrafast probes, and materials synthesis. In this Review, we outline the essential theoretical ingredients underpinning excitonic order and discuss how dimensionality, disorder and screening affect stability. We then examine the diverse experimental fingerprints of the excitonic state, with central focus on strategies to disentangle excitonic order from competing phases such as charge density waves, Mott insulating states, and hybridization-driven insulators, particularly in systems where non-trivial band topology plays a role. We survey the rapidly expanding family of candidate materials, from layered chalcogenides and correlated rare-earth compounds to artificial excitonic platforms and optically driven non-equilibrium condensates. Finally, we discuss the key challenges and emerging opportunities in the field, identifying the theoretical and experimental frontiers that promise to shape the next decade of research.

## Introduction

Excitonic insulators (EIs) are a unique quantum phase of matter in which the spontaneous formation of electron-hole pairs (excitons) drives an insulating state[1-8]. In narrow-band semiconductors and small-overlap semimetals, as the exciton binding energy exceeds the bandgap, holes from the valence band can couple to electrons from the conduction band. This interband Coulomb attraction gives rise to an insulating collective excitonic bound-state, as depicted in Figure 1. Unlike conventional band insulators, whose gaps arise from single-particle band structures[9], or Mott insulators, where strong on-site Coulomb repulsion localizes electrons[10,11], the insulating behavior in EIs emerges purely from many-body electron-hole correlations, forming a macroscopically coherent superfluid condensate characterized by charge neutral excitons[12]. This ordered state is characterized by an order parameter, with collective amplitude and phase excitations, placing EIs, alongside superconductors and density-wave systems, in the framework of macroscopic quantum-coherent phenomena.

Beyond their fundamental definition, EIs provide a natural bridge between semiconductor physics and correlated-electron phenomena, as they offer a unique platform for probing spontaneous symmetry breaking, phase-transition phenomena and strong electron-hole correlations in materials. Theoretically, these systems are predicted to exhibit a Bardeen-Cooper-Schrieffer to Bose-Einstein condensate (BCS-BEC) crossover as a function of the size of the single-particle bandgap ($E_\text{g}$)[6,8], as seen in Fig. 1f: on the semimetal side ($E_\text{g} < 0$), weakly bound e-h pairs undergo a BCS-like transition at $T_\text{c}$, in analogy to superconductors; whereas on the narrow-gap side, strongly bound excitons condense in a BEC-like manner. The study of EIs intersects multiple fields - condensed matter, materials science, and optoelectronics- and has potential applications on excitonic-based transistors, ultrafast optical switches, and quantum coherent devices, making them relevant for a broad scientific community across both fundamental and applied levels.[13–17].

Several decades after excitonic insulators were first proposed[1,2], experimental breakthroughs in the last five years have renewed the relevance of the topic, allowing experimental observation of excitonic signatures in a growing number of candidate materials. These experimental advances include unprecedented resolution in spectroscopic techniques, ultra-fast pump-probe setups and time-resolved photoemission, which now allow direct observation and control of excitonic order and excitation modes on femtosecond timescales. At the same time, progress in material synthesis has seen the emergence of new material platforms beyond the historical candidates (e.g. TiSe$_2$ and Ta$_2$NiSe$_5$), which include quasi-one-dimensional[18] and monolayer crystals[19,20]. On the other hand, advances in theoretical condensed matter physics, particularly within density functional theory (DFT), have provided access to realistic band structures even for complex materials. In parallel, the development of powerful many-body techniques, such as dynamical mean-field theory (DMFT)[21], has enabled quantitatively accurate descriptions of strongly correlated electronic states that go well beyond single-particle frameworks. Building on these developments, predictions have been extended to systems such as graphene-based structures including bilayers[22] and carbon-nanotubes[23], and new two-dimensional classes of materials, such as Ta$_3$X$_8$ (X = I, Br)[24,25]. These developments, combining experimental materials discovery and theoretical guidance, have significantly broadened the scope of materials where excitonic order can be investigated, fueling a surge of high-profile studies that reflect the community's growing interest and motivate the present review.

This review integrates theoretical foundations, experimental fingerprints, and materials platforms providing a comprehensive overview of the state-of-the-art on excitonic insulators. We focus on key developments that have redefined the field over the past decade and on the challenges that will shape the next. Section 1 revisits the essential theoretical ingredients that underlie the exciton insulator state: electron-hole pairing, exciton condensation, and key models (Mott, Kohn, Hubbard, Falicov–Kimball). We revisit the BCS-BEC crossover framework and discuss the effects of dimensionality, disorder, and screening. Section 2 surveys a range of experimental signatures of the excitonic condensate (revealed by techniques such as ARPES, STM, transport, Raman and ultrafast probes), with emphasis on disentangling EI order from possible competing phases like CDW, Mott insulators and topological insulators. Section 3 categorizes the major materials platforms that have been identified so far, including layered chalcogenides, rare-earth systems, and artificial heterostructures, and collects the predictions on candidate materials. Section 4 outlines challenges and opportunities, pointing to along future experimental routes for the realization and probing of EI systems, along with the most promising potential applications.

# 1 Theoretical foundations of excitonic insulators

## 1.1 General considerations

Excitons are the bound states of electron-hole pairs characterized by exciton binding energies $E_b$. In the Wannier-Mott formalism,

$$E_b = \frac{\mu e^4}{2\hbar^2 \varepsilon^2 n^2}$$

where $e$ is the electron charge, $\mu$ is the reduced effective mass, $\varepsilon$ is the dielectric constant of the material, and $n$ is the principal quantum number (the ground state is $n = 1$). Conceptually similar to Cooper pairs, their formation is driven by the interband Coulomb interaction rather than electron-phonon interaction.

Excitons most commonly form out of equilibrium in materials with a band gap $E_g$ after an electron is excited from the valence to the conduction band, leaving a hole behind. In the 1960s, it was postulated that excitons in semimetals[1] and semiconductors[2] behave like bosons and can spontaneously form and condense without external stimuli. These postulates were put on a firmer footing by a series of works[3–5,7,8,12], which characterized the excitonic order by introducing an excitonic order parameter $\Delta = |\Delta|e^{i\varphi}$. This complex quantity measures the spontaneous coherent expectation value $\langle c^\dagger_{c,\mathbf{k}} c_{v,\mathbf{k}} \rangle$ of electron–hole pairs and emerges at a critical temperature $T_c$ (here $c^\dagger_{c/v}$ are electron creation operators in the conduction and valence bands).

Often, this coincides with the opening of a gap at the Fermi level, in which case the state is called an *excitonic insulator*. If the conduction and valence band extrema exist at the same Brillouin zone momentum (direct gap, Fig. 1 **a, c**), the excitons form with zero center-of-mass momentum **Q**, and the order parameter is spatially uniform. If, however, the electron and hole pockets are separated in momentum space (indirect gap, Fig. 1 **d, e**), the electron-hole pairs acquire a non-zero **Q**. This leads to a spatial modulation of the excitonic order parameter, resulting in accompanying states such as charge- or spin-density waves.

Mean-field calculations established that the excitonic phase exists in an asymmetric dome in band-gap-dependent phase diagrams in the vicinity of semimetal-semiconductor transition. These excitonic phases are degenerate between charge-density, spin-density, charge-current, and spin-current density waves when only the density-density part of the Coulomb interaction is taken into account. However, interband scattering lifts this degeneracy, and in the simplest models the spin density wave is the lowest energy state - unless strong coupling to phonons is present, promoting the charge density wave.

These results are limited to the weak-coupling regime, which typically requires significant electron–hole pocket overlap to produce an instability. Moreover, mean-field methods neglect dimensionality effects and can predict long-range order where none exists. Going beyond the BCS-like regime thus required simplified models that nonetheless enable a more detailed treatment of strong interactions.

## 1.2 Strong-coupling regime

The simplest such model is the extended Falicov-Kimball model (EFKM, see Box 5). At half-filling in the strong coupling regime ($U' \gg t_a, t_b$) this model can be mapped to the well-understood spin-1/2 XXZ model with an external magnetic field[26], where spin–spin interactions come from band hopping and the energy difference between orbitals $D$ acts like a magnetic field along the z-axis. This mapping, in which one labels an electron in orbital $a$ as spin up and an electron in orbital $b$ as spin down, enables characterization of the phase diagram. It consists of the band insulator (corresponding to a saturated spin-polarized state), excitonic insulator (the *XY* magnetic order), and staggered orbital order (Néel antiferromagnet along $z$ axis) phases. In one dimension (1D), this analysis can be supported using density matrix renormalization group method that gives the essentially exact ground state phase diagram as a function of $U'$ and $D$ with the excitonic phase understood as a critical phase with power-law correlations[27]. In 2D, the zero-temperature phase diagram exhibits a long-range excitonic order as demonstrated by the constrained path Monte Carlo calculation[28], in good agreement with self-consistent Hartree-Fock results[29]. The Hartree-Fock results also present the same qualitative picture in 3D[29], with excitonic phase surrounding the staggered orbital phase in between the fully band-polarized insulators. However, the EFKM is limited as it includes only spinless electrons, so it does not correctly describe the possible spin-triplet excitonic orders and competition of excitonic phase with magnetic orders. Moreover, as it only includes the onsite interband interaction, it neglects the important effects of the long range nature of Coulomb interaction and associated screening effects.

To overcome the limitations of EFKM, one can include the spin degree of freedom and obtain the two band Hubbard model (2BHM, see Box 5). Its correlated phases arise from the interplay between orbital energy splitting $D$, on-site Coulomb interaction $U$ and Hund's coupling $J$. At half-filling, if orbital splitting dominates over the Coulomb interaction, the system becomes a trivial band insulator (low-spin state), whereas in the opposite limit double occupation of a single orbital site is energetically unfavorable, thus driving the system into the Mott insulator phase (high-spin state). Between these two limits, an excitonic order can potentially emerge, as evidenced by multiple numerical studies[30–39]. Unlike EFKM, the excitonic order parameter in the 2BHM can be decomposed into spin-singlet and spin-triplet components, and their competition can be studied. In the absence of Hund's coupling and pair hopping, these order parameters are degenerate, but when $J > 0$, the spin triplet order parameter is favored, with non-zero pair hopping further supporting the triplet condensate[40]. Similarly to EFKM, deeper insight can be revealed by mapping the Hamiltonian to effective spin models. To instead stabilize the singlet order parameter, either negative Hund's exchange[41] or a strong coupling between the electrons and the lattice is necessary[42]. Although 2BHM improves on EFKM, it remains oversimplified for realistic materials, as it includes only local interactions and thus inadequately describes extended excitons. Unfortunately, the high computational cost limits the reliable treatment of more complex strong-coupling models beyond the Hartree–Fock level. For a more detailed exposition of the strong-coupling results, we refer the reader to excellent reviews by Kuneš[43], Kaneko and Ohta[44].

## 1.3 BCS-BEC crossover

Excitonic order can be viewed from two limits: BCS-like, where electron-hole pairing is a Fermi-surface instability in the semimetallic regime, and BEC-like, where pre-formed tightly bound excitons in the semiconducting regime undergo condensation. Therefore, the existence of excitonic pairs in the strong coupling regime, governed by a separate temperature scale $T^*$, does not imply excitonic order, analogous to preformed Cooper pairs near a superconducting transition. The transition between these limits, controlled by the interaction strength or the band gap, is known as the BEC-BCS crossover[14,45] (Fig. 1 f) and has been studied in continuum[46], lattice EFKM[27,47–50] and 2BHM[51,52]. In the semimetallic BCS regime, Coulomb interactions are screened by finite charge density, yielding loosely bound excitons that are not the lowest energy excitations and emerge only at $T_c$ - so pairing and coherence occur simultaneously. This phase is characterized by anomalous excitation spectra peaked at the Fermi wavevector $\mathbf{k}_F$ and a gap that is directly correlated with the magnitude of the order parameter. By contrast, in the semiconducting BEC regime, weaker screening allows excitons to exist without long-range phase-coherence as well-defined bosonic quasiparticles within the energy gap below the particle-hole continuum at $T^* > T > T_c$. Here, the anomalous excitation spectra are widely spread in momentum space, and the energy scales of the single-particle gap and the order parameter become different. The crossover also affects collective modes (see section 2.2): both the amplitude and phase mode give prominent spectral peaks in the BCS regime, while the amplitude mode becomes strongly suppressed in the BEC regime[53].

## 1.4 Effects of screening, dimensionality, and disorder

As the excitonic order relies on interband Coulomb interaction between electrons and holes, dielectric screening and dimensionality of the system strongly affect pairing and ordering properties. In general, reduced dimensionality leads to weaker screening and larger excitonic binding energies, increasing from tens of meV in three-dimensional materials, to hundreds of meV in two-dimensional layers, to several eV in 1D and quasi-1D systems. Screening also depends on carrier density, leading to the appearance of a dome at the semimetal-semiconductor transition. The sensitivity to screening often also necessitates going beyond the static screening approximation and considering dynamic screening, which can strongly renormalize pairing strength and transition temperatures[54].

However, reduced dimensionality also results in increased long-range fluctuations, reducing the tendency of excitons to order. While at the mean-field level the condensation occurs in every dimension, true long-range order occurs at finite temperature in three dimensions, and only at $T = 0$ in two dimensions. However, the onset of excitonic order occurs when power-law correlations in the ground state appear[27] which in two-dimensions at finite temperatures has the nature of Berezinskii-Kosterlitz-Thouless (BKT) transition[55].

Disorder also affects exciton formation, as impurities have a pair-breaking effect[56], analogous to magnetic impurities in superconductors. They affect the magnitude of the order parameter, the transition temperature, and density of states, with a critical impurity density above which the excitonic order vanishes. Impurities can also result in bound states in the excitonic gap[57-59]. Disorder influences the transport of the collective modes of the system, depending on whether the scattering mixes electrons and holes: If the impurity potential preserves symmetry, the collective modes are robust to disorder and propagate ballistically, affecting low-temperature thermal transport[60]. In systems where excitonic and superconducting order compete, disorder generally promotes an instability to superconductivity, but it can also enable a regime where both orders coexist[61].

## 2 Experimental Fingerprints of Excitonic Insulators

Establishing the EI phase experimentally relies on identifying measurable fingerprints of electron-hole condensation. As excitonic order often intertwines with other mechanisms such as lattice distortions, charge-density waves, or band insulators, no single observable provides a definitive signature. In this Section, we review the most important signatures of exciton order along with their current state-of-the-art interpretations and discuss excitonic interaction with its main competing phases. The key results are collected in Table 1. The view-point of this part is mainly set on canonical pristine layered compounds, which give a solid template for identifying general excitonic fingerprints. As this remains a rapidly evolving topic, these fingerprints should be considered indicative rather than conclusive.

### 2.1 Band reconstruction

#### 2.1.1 Indirect vs. direct gap systems

In indirect-gap systems, exciton order breaks translational symmetry producing a structural reconstruction which multiplies the original unit cell, resulting in a smaller Brillouin zone size (Fig. 1d,e). This leads to band folding (or *backfolding*), with replicas of the original bands appearing at high-symmetry points of the first BZ (Fig. 2a). The backfolding-induced band hybridization results in a redistribution of spectral weight[62-69] and metal–insulator transition with (temperature dependent) gap opening[63,65-69], often accompanied by valence-band flattening and a shift to higher binding energy. These effects arise from hybridization-induced electron-hole coupling[63,67]. Some excitonic candidates show backfolding only of the topmost VB[69] (see Section 3.1.1), a potential smoking-gun of excitonically-driven band reconstruction. Backfolding of the VB together with spectral weight redistribution and top-band flattening, have surprisingly been observed slightly above $T_c$ in some cases[62,65-67]. This anomalous temperature dependence can be interpreted as strong deviation from a conventional periodic lattice distortion (PLD) driven phase-transition[67]. Time-resolved ARPES further shows that the EI-formation associated spectral weight collapses, upon pump-pulse, on ultrafast (∼20 fs) timescales[64], characteristic of electronic dynamics, faster than phononic ones.

For direct-gap EIs, where no translational symmetry is broken, CDW and band-folding are not observed. The VB top can however exhibit pronounced flattening[70] and shift toward higher binding energies[71] as the temperature decreases and the excitonic gap opens[71,72], an effect also observed via STS[73] and optical conductivity measurements[74-76]. Interestingly, also in this scenario, non-zero band flattening has been observed above $T_c$ suggesting strong exciton-order fluctuations[71,72], with tr-ARPES experiments revealing deformation of the band structure[77] in the ultrafast -electronic- timescale.

#### 2.1.2 Topological excitonic insulators

The few materials proposed as topological excitonic insulators - a system hosting topologically protected edge states in which the insulating gap is opened by spontaneous exciton condensation - have not shown, so far, evidence of backfolding, regardless of their indirect-gap character. Nevertheless they have shown spontaneous gap opening at low temperatures[18,78-80], accompanied, in some cases, by the appearance of in-gap flat bands[18,78], analogous to impurity states in superconductors[78], invoking a BCS-like order. Additional experimental topological EI candidates would be required to determine whether these behaviors are material-dependent or a general feature of this family.

### 2.2 Collective modes

The exciton condensate can be described at the mean-field level by a complex order parameter $\Delta = |\Delta|e^{i\varphi}$. The formation of bound electron-hole pairs supports collective excitations associated with fluctuations of both the amplitude and phase of the order parameter[81]. Following the framework developed for collective excitations in superconductivity, cold atom lattices and

CDWs[82–84], the modes can be classified as gapped (Higgs) amplitude modes, associated with the fluctuation of the magnitude of the exciton, and and gapless (Nambu-Goldstone or Anderson–Bogoliubov) modes, associated with fluctuations of the phase $\varphi$. A detailed theoretical description of the expected excitation structure has already been covered in previous works[44]. Here, we note that different experimental techniques preferentially couple to amplitude, phase, or mixed modes, depending on whether they perturb the magnitude, phase or associated electronic density of the order parameter.

### 2.2.1 Amplitude modes

The Raman vertex (i.e., the operator that describes how the electronic degrees of freedom couple to photons) transforms as a fully symmetric representation, efficiently coupling to the fully symmetric amplitude fluctuations of the order parameter. As a result, Higgs modes are expected to appear as a prominent or moderately broadened low-energy feature in the Raman response[85–92]. Complications arise in case of linear coupling between the excitonic order parameter and certain lattice distortions, such as the $B_{2g}$ phonon in Ta$_2$NiSe$_5$. This interaction mixes electronic and lattice degrees of freedom, leading to broadened phonon lineshapes and asymmetric Raman profiles near the condensation/structural transition temperature.

### 2.2.2 Phase modes

Pure phase fluctuations are generally Raman-inactive but accessible via optical and THz spectroscopy, as they couple to the field in the presence of a finite interband dipole matrix element or intraband charge motion[53,83]. The THz response of an EI is expected to exhibit a low-energy peak associated with the phase mode, gapped by the generally present interband hybridization and coupling to lattice distortions, and accompanied by a redistribution of spectral weight as the excitonic gap forms. Experimental THz and optical studies, however, suffer from the fact that strong exciton–phonon coupling can hybridize these phase modes with infrared-active phonons and can complicate their identification[88,89].

Ultrafast spectroscopy has emerged as a key tool to distinguish excitonic contributions from lattice ones, due to the differences in typical electron and lattice timescales. Mean-field calculations show that a pump pulse can transiently suppress or enhance the condensate, depending on the excitation frequency and its BEC or BCS nature, with relaxation accompanied by coherent excitonic–phononic oscillations[93]. Nonlinear theory further predicts that it is possible to resonantly excite a massive phase mode, generating a characteristic second-harmonic in-gap signal via parametric amplification[83]. In experiments, ultrafast non-linear excitations and time resolved reflectivity have shed light on the amplitude modes[88,94], while ultrafast microscopy has revealed a coherent, micrometer-scale propagation of collective modes, a phenomenon attributed to the hybridization between the phonon modes and the excitonic phase mode[95] (Fig. 2d).

## 2.3 Lattice effects

Charge density waves (CDWs), typically accompanied by a periodic lattice distortion (PLD) resulting from lattice instabilities[96–98], have traditionally been interpreted within the Peierls framework, where their formation is associated with complete or partial Fermi-surface nesting[99]. In EIs, they can coexist with lattice effects such as PLD, electron–phonon coupling[100,101] and phonon softening. It is then of great importance to disentangle excitonic-driven phenomena with lattice distortion as a by-product from lattice-mediated transitions[63,102].

### 2.3.1 Charge density waves vs. periodic lattice distortion

In indirect-gap EIs, the nonzero **Q** vector coupling electron and hole pockets gives rise to CDWs (see Section 1.1) below $T_c$[69,103–106] (Fig. 1d,e), as seen in FT-STM[66,67] (see Fig. 2e). Interestingly, while associated PLD has been reported in certain materials[68,103,107], it does not systematically occur in all indirect-gap systems[68,69]. Very recently, CDW-like DOS modulation imprinted on the edge state of a candidate topological insulator has been observed and linked to SDW/CDW formation[80], although so far structural measurements have not detected PLD[19] (see Section 3.1.2).

In the direct-gap and Dirac-point scenarios, other lattice distortion effects can coexist with the excitonic ordering, like unit-cell preserving[74,86,108] structural transitions[109,110] or exceptionally small lattice distortions. Material platforms where small or absent lattice distortions cannot account for the significant spectroscopic renormalization[18,74,108,111], point in the direction of exciton-formation driven transitions.

### 2.3.2 Phonon-softening

Phonon softening (blueshift) constitutes a complementary fingerprint of excitonic order when it is directly correlated with electronic instabilities. Raman spectroscopy, inelastic x-ray scattering, and diffraction experiments near the transition temperature can reveal softening and broadening of specific phonon modes, helping to distinguish between electron–phonon coupling intertwined with excitonic fluctuations and purely structural instabilities[86,90,91,112–114].

The softening of an optical phonon branch, established by x-ray scattering[100,115] or Raman[116,117], when combined with evidence for electronic band renormalization and sub-gap collective modes, supports an interpretation in terms of exciton–phonon coupled order rather than a purely lattice-driven CDW transition (see Fig. 2b and Section 3.1.1 for further details).

## 2.4 Order melting

Tuning the occupation of the valence and conduction bands, via chemical doping, electrostatic gating, pressure, or photoinduced carrier injection, provides a controlled route to melt the excitonic order. Tracking how the system departs from the ordered phase under such perturbations offers valuable fingerprints of this excitonic many-body phase.

### 2.4.1 Chemical and electrostatic doping

Chemical doping modifies the carrier population by shifting the chemical potential or altering the band structure. In both cases, the resulting increase in carrier density, which alters the screening (see section 1.4), tends to occur at the expense of excitonic order[69,78,103,118–121] (see Fig. 2c and Sections 3.1.1, 3.1.2). Excitonic order suppression can also be achieved by electrostatic gating[79,80]. Nevertheless, when the induced charge density is low enough that Coulomb screening is negligible, other effects—in particular those favoring robustness of EI—may become significant. This is the case of $Ta_2NiSe_5$[122], where slight n-type doping changes the normal-phase band structure in such a way that the bandgap is reduced, strengthening e-h pairing.

### 2.4.2 Pressure

Increasing pressure generally causes the valence and conduction bands to shift rigidly toward increased overlap (i.e., towards the semimetallic state), thus pushing the Fermi level deeper into both bands and increasing Coulomb screening. This can dislodge the excitonic phase from its window of stability as proposed by Halperin and Rice[8], thereby providing a characteristic fingerprint of EI under pressure: a pressure-induced melting of the ordered phase[74,119,120,123,124] (see Fig. 2c and Section 3.1.1).

## 2.5 Transport anomalies

Electronic transport provides a window in the formation of the excitonic states, as the formation of bound electron-hole pairs suppresses single-particle transport while preserving charge neutrality. Transport measurements carry information not only on the presence of an energy gap, but also about its symmetry with respect to charge neutrality and its collective origin.

### 2.5.1 Resistivity and Hall anomalies

In correlated electron-hole bilayers (see Section 3.3.2) evidence for exciton condensation has been demonstrated by measuring the Coulomb drag[125,126]. In pure-compound EIs, the opening of the excitonic gap $\Delta_{EI}$ is directly observed as a drop in conductivity at $T_c$[19,20,74,127,128] (see Sections 3.1.1, 3.1.2), causing the collapse of the number of free carriers, which in turn leads to a progressive loss of the Drude weight as $T_c$ is approached, as measured by optical conductivity.

The Hall effect furnishes a particularly discriminating test. Because excitonic pairing condenses the same number of electrons and holes, the Hall carrier density measured below $T_c$ is expected to drop markedly[20,129] (see Fig. 2f). Thus, concurrent observation of a collapsing Drude weight in optical conductivity coupled with a drop in Hall carrier density provides strong transport evidence for charge-neutral pairing rather than carrier localization or selective gap opening. Another key feature of EI is that, up to a threshold charge density, the system remains gapped to charge transport, even upon addition of charge (which can be measured capacitively)[19].

In 2D EIs, reduced dimensionality and screening from the substrate heavily affect these transport responses. Reduced screening enhances exciton binding (see Section 1.4) and helps stabilize EI order over larger parameter ranges; transport in the 2D limit can therefore approach a charge-neutral insulating state at compensation, with vanishing Hall response and strongly suppressed longitudinal conductivity[19,20]. Electrostatic gating provides a direct experimental knob: introducing carriers can induce an ambipolar collapse of the excitonic bandgap, enabling a transport-based test of the pairing hypothesis, as indicated by recent monolayer studies[79].

### 2.5.2 Magnetotransport

Strong magnetic fields provide a tuning parameter through both orbital quantization and Zeeman coupling, which can influence the excitonic binding energy and the coherence of the condensate, and can both destabilize or stabilize excitonic order depending on coupling strength and bandstructure (see section 3.2). Quantum oscillations, a phenomenon typically observed on metals in a magnetic field, have recently been reported in excitonic insulators[130]—where a Landau fully quantized regime is observed above a critical field—. While care must be taken in interpreting these oscillations, which can arise in similar devices due to the graphite gate used in the experiments[131], more recent works have provided more direct evidence that quantum oscillations can arise in dipolar exciton condensates. In particular, two recent studies report oscillatory magnetotransport consistent with neutral, yet dipole-carrying, excitonic quasiparticles[125,126]. Further investigations are required to extrapolate more universal trends.

## 2.6 Disentangling excitonic order from competing phases

Not only is it difficult to distinguish between different metal-insulator transitions — namely Peierls, excitonic, Mott and Kondo (Figure 3) — but the situation is further complicated by the possible competition and coexistence of these exotic phases, either simultaneously or within the same phase space (*e.g.* PLD-induced Mott insulator 1T-TaS$_2$[132,133], topological Kondo insulator SmB$_6$[134–138], topological exciton insulator ML 1T'-WTe$_2$[19,79,80,139]). Correlated systems often exhibit intertwined orders, possibly with the same symmetry, making it difficult to determine a leading interaction (if one exists). The contentious exchange in the last decade surrounding the gap opening mechanism in Ta$_2$NiSe$_5$[53,72–77,85–90,93,94,112–114,121,122,140–150], for example, illustrates that while this process is difficult and not entirely conclusive, much progress has been made in developing improved theoretical descriptions and identifying experimental signatures.

tr-ARPES presents strong *quantitative* distinction between the phases, differentiating the order melting times of structural CDWs (slow atomic lattice timescale, ~ 60–300 fs) and electronic phases (< 100 fs)[151]. These are then further separated, with the fast melting of Mott phases governed by short hopping times (typically < 10 fs), while e-h recombination exhibits moderate timescales and a unique dependence on induced carrier density (measured 77 ± 9 and 44 ± 14 fs for energy densities 300 and 600 Jcm$^{-3}$ in 1T-TiSe$_2$[133], respectively). Similar information can be extracted from time-resolved optical reflectivity measurements[132]. Care must be taken when comparing the effects of pump energy density on relaxation times between materials as the size of the Fermi surface will affect the concentration of induced carriers. Additionally, for complex systems with may physical mechanisms leading to formation of the EI state, tr-ARPES potentially reveals only the dominant physics at equilibrium, and complementary evidence from other sources may be required to construct a complete physical understanding.

STM provides highly complementary local information, capable of measuring the spatial dependence of order parameters, as well as exploring electronic and magnetic field phase spaces which are commonly inaccessible to photoemission studies. For instance, the Kondo insulator can be identified by the magnetic field dependence of its band gap[152], while the exciton insulator quasiparticle gap should remain relatively insensitive to magnetic field. The superior energy resolution allows for the analysis of in-gap states (exciton wavefuction decay length in Ta$_2$NiSe$_5$[144], excitonic Yu-Shiba-Rusinov states in Ta$_2$Pd$_3$Te$_5$[153], CDW-order in the 1T'-WTe$_2$ topological edge[80]) albeit without momentum resolution. In the case of a CDW order parameter, as the degree of electron-phonon coupling and therefore PLD that is expected for the EI phase is still inconclusive, STM and similar metrology techniques (TEM, x-ray diffraction) as well as phonon softening in Raman spectroscopy, while still producing essential supporting evidence, do not provide the same level of distinction as time-resolved techniques.

## 3 Where Theory Meets Reality: Materials and Platforms

Realizing the EI phase experimentally requires materials that combine small band gaps or semimetallic overlap with weak screening, allowing Coulomb attraction to dominate. Such conditions arise only in finely balanced systems, often near competing structural or electronic instabilities. Over time, research focus has expanded from hunting for candidate platforms in particular materials classes to also artificially engineering the conditions for EIs, as in heterostructures and light-driven non-equilibrium states. The following subsections trace this progression, illustrating how diverse material systems realize the same fundamental excitonic mechanism under different microscopic conditions.

### 3.1 Layered chalcogenides

Layered chalcogenides constitute the foremost platform for the realization of EI-based devices (see Sect. 4), and provide the foundational case studies that established the key experimental fingerprints discussed in Section 2. Crucially, their structural simplicity, combined with their sensitivity to interlayer coupling, has positioned them as ideal platforms for dimensional engineering. Reducing these materials to the monolayer limit (Section 3.1.2) or assembling them into van der Waals heterostructures (Section 3.3.2) provides control over the parameters governing excitonic order.

#### 3.1.1 Bulk layered chalcogenides: canonical case studies

Bulk 1T-TiSe$_2$ is a prototypical candidate for a cooperative EI[111,154,155], where electronic and lattice instabilities are strongly coupled. Its transition to an ordered state at $T_c \approx 200$K[103,156,157] is marked by several key experimental fingerprints discussed in Section 2. Band reconstruction, a hallmark of indirect-gap condensation, has been observed via backfolding of bands and gap opening in ARPES[62,158] (see Fig. 3a and Section 2.1). The electronic origin of this transition is further underscored by ultrafast pump-probe studies[64,133,159], which reveal the melting of the ordered state on the femtosecond scales (see Section 2.4). Meanwhile, this electronic order coexists with a lattice distortion, leading to a pronounced softening of optical phonons and the appearance of a collective mode in Raman spectroscopy (see Sections 2.2, 2.3). The intertwined nature is further highlighted by the existence of a hybrid phonon-excitonic mode[100,160,161] (Fig. 4a) and the emergence of superconductivity under pressure[123] or upon chemical doping[118], suggesting a finely balanced ground state where excitonic and lattice instabilities are mutually reinforcing. Despite decades of study, the precise role of the lattice versus purely electronic correlations remains an open question[102].

Ta$_2$NiSe$_5$ provides a complementary example of a cooperative EI. As a direct-gap system[70–72,149], it exhibits pronounced flattening of the topmost valence band (Fig. 4b)—a key fingerprint of the excitonic transition (see Section 2.1). The correlation-driven nature of this state is further confirmed by the collapse of the insulating gap at low temperature upon doping[121,122], under pressure[74,124], as well as by ultrafast order melting studies[140,141,148,162] (see Section 2.4). Supporting this picture, Fukutani et al.[142] report characteristic excitonic signatures above $T_c$, providing evidence for preformed excitons that stabilize the low-temperature EI phase. Concurrently, the accompanying lattice symmetry lowering[90,109,112] and phonon softening[113,141,147] (see Section 2.3) across the transition highlight the intertwined nature of the excitonic and lattice degrees of freedom.

$Ta_2Pd_3Te_5$ is an example of EIs driven by a predominantly electronic instability, as across its semimetal-to-insulator transition[18,78,94,163–166], the material exhibits negligible structural changes. ARPES measurements on bulk crystals reveal a temperature-dependent band gap opening (Fig. 4c), signifying a transition from a high-temperature Dirac semimetal to a low-temperature insulating state[18,78,165,166], consistent with the exciton condensate (see Section 2.1). More recently, optical conductivity signatures[167] have provided direct evidence of the condensation, showing the collapse of the Drude response (Section 2.5.1) and the appearance of exceptionally narrow absorption peaks at low temperatures. Its predicted 2D nontrivial band topology[168,169] has been corroborated by ARPES and STS measurements[163,165], positioning $Ta_2Pd_3Te_5$ as a promising candidate for a topological EI (Section 2.1), where excitonic order and protected boundary states coexist.

### *3.1.2 Dimensional reduction and monolayer platforms*

Dimensionality reduction boosts many-body interactions while simplifying the structural landscape, thus creating an ideal setting for stabilizing excitonic order. This paradigm is exemplified by monolayer 1T-$TiSe_2$, where reduced interlayer coupling and dielectric screening increase the CDW transition temperature from $\approx 200$K to $\approx 230$K[65,97,170]. While some works attribute the origin of the microscopic order to Peierls-like lattice distortion[171,172], an excitonic instability[173] scenario is strongly supported by the observation of temperature-dependent backfolded electronic bands in the CDW state via high-resolution ARPES[66]. Furthermore, the suppression of the ordered phase by modest doping[174,175] or defect introduction[176] underscores its correlated electronic nature.

Enhanced interactions in the 2D limit also stabilize an EI phase in MBE-grown monolayers of 1T-$ZrTe_2$[67,68], which exhibit temperature- and doping-dependent gap formation in ARPES. In 1T-$ZrTe_2$ the band renormalization occurs above $T_c$, being attributed to the preformed excitons scenario[67], analogous to the fluctuations observed in bulk $Ta_2NiSe_5$ (Section 2.1). Similarly, monolayer $HfTe_2$, an indirect-gap EI candidate but without PLD, exhibits backfolding of the topmost valence band in the excitonic state[69], a signature also seen in bulk $TiSe_2$. Here, electron doping progressively suppresses $T_c$, ultimately melting the excitonic order beyond a critical carrier density.

Monolayer 1T'-$WTe_2$ represents a distinct example of dimensional reduction stabilizing coexisting quantum phases. In the atomically thin limit, this correlated topological semimetal[177] exhibits a gapped bulk state, whereas the transport signatures[19,20]—a V-shaped dependence of conductivity on carrier density and a distinct chemical potential step at charge neutrality—point to simple single-particle insulator but consistent with equilibrium exciton formation (see Section 2.5). Crucially, electrostatic gating provides direct access to the condensate, driving an ambipolar quantum phase transition[79,80], in which the quasiparticle gap collapses abruptly upon doping (see Fig. 5b and Section 2.4). The absence of CDW signatures in the 2D bulk[19] points to a primary electronic, possibly spin-orbital entangled ground state[80,178] which manifests as CDW/SDW imprinted on the edge state[80]. Notably, the emergence of superconductivity under electrostatic doping[179,180] raises the possibility of exciton-mediated pairing[181,182], positioning monolayer $WTe_2$ as a uniquely tunable platform where topology, excitonic condensation, and unconventional superconductivity converge.

## 3.2 Rare-earth and mixed-valence systems

Rare-earth and mixed-valence compounds[183] form a distinct class of EI systems, where the instability arises from hybridization between localized $4f$ and itinerant $5d$ states. The resulting charge-transfer excitons naturally intertwine electronic and magnetic correlations, providing a fertile platform for exploring the interplay between excitonic condensation and magnetism.

In the mixed-valence compound $TmSe_{0.45}Te_{0.55}$[184] excitons form between localized $Tm^{2+}$ $4f$ holes and itinerant $Tm^{3+}$ $5d$ electrons. At ambient pressure, the compound is a narrow-gap semiconductor ($E_g \sim 0.15$ eV) without long-range order. Under pressure, however, A semiconductor-to-insulator transition below $\sim 250$ K was identified, and interpreted as the onset of strong excitonic correlations, and a linear rise in thermal diffusivity below 20 K[185,186] (see Fig. 4d), often associated with exciton condensation or superfluid-like behavior[127,184,186,187]. Theoretical studies[188,189] both support and complicate this interpretation, leaving open whether the pressure-driven anomalies reflect excitonic ordering or band reconstruction.

$SmB_6$, the archetypal topological Kondo insulator[134,135], exemplifies how strong correlations and topology can intertwine with excitonic physics. While it develops a hybridization gap from Sm $4f$-$5d$ interactions and hosts topologically protected surface states, its low-temperature bulk properties remain enigmatic[190]—exhibiting a metallic-like specific heat[191,192], residual thermal conductivity[192], and quantum oscillations[193,194] deep within the insulating regime. An excitonic scenario[195] offers a compelling resolution: the "Mexican hat" band structure, as shown in Fig. 4e, is susceptible to neutral excitons and magnetoexcitons, explaining these anomalies. Muon spin rotation experiments[196] detect the gradual freezing of a low-energy ($\sim 1$ meV) bulk spin exciton below 20 K, coinciding with surface conductivity saturation, indicating a key role of spin-exciton scattering in linking bulk correlations to surface transport.

Bilayer Sr$_3$Ir$_2$O$_7$[197,198] has emerged as a host for an antiferromagnetic EI state[8], where spin–orbit and strong electronic correlations underpin its insulating behavior. Beyond its known collinear antiferromagnetic order below $T_N \approx 285$ K[197,198], recent theory and experimental works[199-201] have revealed features (Fig. 4f) suggestive of excitonic ordering intertwined with magnetism, motivating further work to clarify the microscopic origin of this proposed antiferromagnetic EI phase.

### 3.3 Artificial excitonic platforms

Artificially engineered quantum structures—such as semiconductor quantum wells and van der Waals heterostructures—offer additional control over electronic band alignment and carrier confinement. In parallel, advances in ultrafast optical spectroscopy have opened a new frontier for probing transient, non-equilibrium excitonic phases, revealing emergent collective behaviors inaccessible in equilibrium.

#### *3.3.1 Semiconductor double quantum wells*

Coupled quantum wells (Fig. 6a) are widely studied as ideal semiconductor BEC platforms due to their precise band-engineering capabilities. In a typical coupled quantum well system[202], two semiconducting quantum well layers are separated by a thin insulating barrier. A perpendicular electric field tilts the band structure to bring conduction band minimum and valence band maximum in different quantum wells, leading to the spatial separation of electrons and holes and facilitating the formation of indirect exciton states. As a result, photoexcited excitons can propagate over macroscopic distances (e.g., >10 μm) before recombination, which is observed as a photoluminescence ring around the excitation origin spot, observed in GaAs/AlGaAs coupled quantum wells[203–206]. Furthermore, within the condensate, electrons and holes are tightly bound. Therefore, an electron current in one well drives an equal hole current in the other, resulting in a perfect Coulomb drag effect as reported by Nandi et al.[207].

Electron-hole bilayer structures can also be formed in in double quantum well systems where the conduction and valence bands in adjacent wells have a finite spatial overlap. Such engineered band structures can spontaneously give rise to equilibrium excitons analogous to the photoexcited coupled quantum well systems but without external excitation. The InAs/GaSb quantum well[208] is a notable example, exhibiting a unique inverted band structure where electrons reside in the InAs layer and holes in the GaSb layer.

#### *3.3.2 vdW heterostructures*

The principle of realizing exciton condensates in spatially separated electron-hole bilayers has been successfully extended to vdW heterostructures (Figs. 6b and c), a leap enabled by the recent progress in exfoliation and transfer techniques. The rich vdW material family—including semimetallic graphene, semiconducting transition metal dichalcogenides (TMDs, e.g., MoS$_2$, WSe$_2$), and insulating hexagonal boron nitride (hBN)—thus provides an exceptional and versatile toolbox for constructing tailored electron-hole bilayer systems with atomic precisions. Examples of such engineered systems include graphene double layers[209–212], TMD double layers[213–216], which are electrically isolated by ultrathin hBN spacers. Moreover, moiré potentials, arising from relative twist or lattice mismatch between the layers, provide a novel quantum knob. These potentials can flatten electronic bands, enhance correlations, and are predicted to boost the critical temperature for condensation. For more detailed discussions, see the recent review by Moon et al.[217].

#### *3.3.3 Non equilibrium EI*

Transient non-equilibrium condensates can be induced by ultrafast optical pumping even in systems whose equilibrium ground states are semimetallic[218,219] or band insulating[220,221]. A key distinction from the equilibrium EI, first emphasized by Östreich and Schönhammer[222], is that the non-equilibrium EI are characterized by self-sustained oscillations in the macroscopic polarization with a finite, time-dependent amplitude, whose frequency is tunable via the absorbed energy. A concrete realization of this scenario was predicted in Dirac materials[218,219], in which photoexcitation can induce an excitonic gap simultaneously both in the electron and hole bands - signature of a transient exciton condensate. The predicted gap magnitude (10–100 meV) is experimentally accessible via techniques such as time-resolved ARPES. Notably, such a transient condensate has recently been directly observed in the surface states of topological insulator Bi$_2$Te$_3$[223] (see Fig. 6d).

### 3.4 Beyond canonical candidates

The search for the excitonic insulator state has expanded to materials beyond the extensively investigated layered chalcogenide template as discussed in Secs 2, 3, offering fundamentally different mechanism or order parameter symmetries. Two distinct frontiers illustrate such expansion: the long-sought realization of an excitonic gap in graphene—a model Dirac system, and the prediction of a novel spin-triplet condensate in a 2D ferromagnet such as Ta$_3$X$_8$ (X = I, Br).

Graphene has long been viewed as an archetype for interaction-driven EIs. Its Dirac semimetallic band structure, with a vanishing density of states at the Dirac point, suppresses screening and enhances long-range Coulomb interactions[224,225]. The interaction strength is characterized by the ratio between the Coulomb potential and kinetic energy, i.e., coupling constant $\alpha = e^2/4\pi\varepsilon\varepsilon_0\hbar v_F$[225]. In freestanding graphene ($\varepsilon = 1$), in which $\alpha \approx 2.2$, the unscreened Coulomb interaction has been predicted to result in spontaneous mass generation[226–231] (see Fig. 6e). The resulting low-energy ground state is a coherent macroscopic condensate often referred to as an EI[1,232]. Experimentally, however, this phase remains unobserved: substrate screening reduces $\alpha$ to subcritical values ($\sim$0.8), while even suspended graphene shows no insulating gap in transport measurements down to millikelvin temperatures[233], due to a renormalization of the Fermi velocity.

A distinct theoretical frontier is represented by monolayer $Ta_3X_8$ (X = I, Br) (Fig. 6f), a recently predicted class of spin-triplet EIs[24]. First-principles calculations reveal exceptionally large exciton binding energies (1.5-2.0 eV) that exceed the quasiparticle gaps, driving a robust excitonic instability. These 2D ferromagnetic semiconductors[234,235] feature nearly flat, opposite-spin band edges derived from Ta $d_{z^2}$ orbitals. Crucially, the corresponding interband transitions are parity- and spin-forbidden, strongly suppressing dielectric screening and amplifying electron–hole attraction. The resulting tightly bound Frenkel-like spin-triplet exciton ($S_z$ = 1) is predicted to condense into a state supporting spin supercurrent[236]. The $Ta_3X_8$ family thus offers a unique platform where excitonic order may intertwine with magnetism and spin texture, pointing toward electrically tunable spin-coherent quantum devices.

# 4 Challenges and opportunities

## 4.1 Future experimental routes

Thanks to the advances in the synthesis and ultrafast spectroscopy, the field has broadened, including not only 3D crystals but also 2D materials. The main outstanding challenges lie in the active control and engineering of excitonic condensates, requiring (i) ultrafast, quantum-coherent schemes to manipulate order and collective modes, and (ii) novel material synthesis to realize robust, high-temperature platforms with tailored quantum functionalities. In particular, devices exploiting the macroscopic quantum coherence of the exciton condensate appear promising, but experimental realization remains a challenge.

### 4.1.1 Ultrafast Control and Coherent Excitonic Dynamics

Recent observations of ultrafast order melting and the spectroscopic mapping of collective modes in EI candidates (Secs.2.1, 2.2, 2.6) have shifted the field from static identification toward dynamic interrogation. A central objective is the coherent control of the order parameter—a prerequisite for novel excitonic-based devices, as discussed in Section 4.2. Collective modes (Sec 2.2) provide a natural basis for phase- and amplitude-engineering schemes using tailored multi-pulse THz and mid-infrared fields to resonantly drive collective excitations, manipulate electron–hole phase coherence, and enable all-optical switching on femtosecond timescales. Beyond equilibrium, optically induced non-equilibrium excitonic insulators in Dirac materials (Sec. 3.3.3) establish a pathway to on-demand condensation via ultrafast, momentum-selective photo-doping that bypasses slow thermalization. Time- and angle-resolved spectroscopies will be essential for tracking non-equilibrium BCS–BEC crossover dynamics, while quantum-optical probes—such as photon interferometry and noise measurements in excitonic drag currents—are required to establish macroscopic quantum coherence beyond transport signatures alone.

### 4.1.2 Novel Material Synthesis and Stabilization Strategies

The evolution of material platforms discussed in Sec. 3, shifted the synthesis challenge toward designing systems with enhanced stability, elevated critical temperatures, and new functionalities. For instance, moiré superlattices in twisted vdW heterostructure[217] offer a particularly powerful route, as reduced dielectric screening and flat-band formation can strongly enhance exciton binding and stabilize condensates at higher temperatures. The application of high magnetic fields offers a distinct physical route to stabilize or uncover exotic phases, including correlated quantum Hall–like excitonic liquids[126]. In parallel, a key priority will be the experimental realization of predicted but unrealized exotic platforms. A distinct example is the spin-triplet excitonic insulator, where the condensation of spin-1 excitons is predicted to give rise to a spin-superfluid state, promising for dissipationless spin transport.

## 4.2 Potential applications

A key advantage of equilibrium EIs is that macroscopic coherence emerges without optical pumping or external energy input. As genuine ground states, equilibrium EIs enable steady-state, low-dissipation operation, conceptually analogous to superconductors and topological insulators. Although charge neutral, the phase-coherent condensate can support dissipationless or weakly dissipative transport, motivating proposals for ultra-low-energy switches, excitonic Josephson-like devices, and phase-coherent interlayer elements. Outstanding challenges include disorder-induced pair breaking, inhomogeneous exciton linewidth broadening, decoherence, dielectric screening from the environment, and the need for encapsulation and scalable electrical control.

For a comprehensive overview of excitonic devices based on quantum wells, the reader is directed to Ref.[237].

### 4.2.1 Exciton-based transistors

Advances in the fabrication of TMD heterostructures paved the way for interlayer exciton(IX)- based field effect transistors (FET), as charge separation into different layers substantially increases the lifetime of the exciton (see Sect. 3.3.2), enabling diffusion lengths of up to approximately 5 μm at room temperature.[238] The out-of-plane electric dipole arising from charge confinement can be modulated by external electric fields, providing a means to control exciton flow. hBN-encapsulation improves both the IX lifetime and its linewidth,[239,240] providing the foundation for the realization of TMD-based exciton-based transistors.

Different approaches for the realization of exciton transistors involve controlling the formation or coherence of the excitons, rather than their flow. One such example is based on the prediction of an excitonic Josephson effect.[241] This involves the formation of a Josephson-like tunnelling current between two spatially separated electron/hole layers. In the "on" state, a phase-coherent excitonic wavefunction is established across the device, enabling a dissipationless interlayer supercurrent. The application of a gate voltage to one of the layers would break phase coherence, controlling the interlayer current and bringing the device in the "off" state. Future proposals may also leverage on different mechanisms to toggle "on/off" the condensate, such as demonstrated excitonic gap collapse for equilibrium EI[79], the proposed transition from band insulator to superfluid exciton condensate in bilayer graphene under a perpendicular electric field[22] or the higher resistance predicted at EI-semimetallic junctions - distinct from the one observed at a semimetal-semiconductor junction[242], which could correspond to the "on" and "off" states for the device.

### *4.2.2 Excitonic platforms for quantum information*

Josephson-like interlayer tunnelling[243], (see Sect. 4.2.1) could, in principle, also enable qubit implementations based on controlled phase differences or collective excitations of the condensate. Although implementations remain at an early stage, recent progress has enabled longer exciton lifetimes , enhanced coherence through spatial superfluidity and correlated insulating states[244], and improved scalability in layered moiré systems like $MoSe_2/WSe_2$ and $WSe_2/WS_2/WSe_2$[44,245]. At the same time, the development of cleaner interfaces and improved electrostatic control enhances prospects for realizing robust nonlinear elements analogous to Josephson junctions. Ultrafast optical control further enables access to transient condensate states[95,140], opening possibilities for switchable quantum coherence[77,79], collective excitations and photon correlations through light-matter entanglement in cavity-coupled systems[246].

## 5 Conclusions

Excitonic insulators have progressed from long-standing theoretical proposals to experimentally verified quantum platforms, supported by converging evidence across synthesis methods, experimental probes techniques, and advances in theoretical modelling. Combining evidence across different material famlies, we explored how Coulomb interaction, lattice distortions, topology, and dimensionality effets cooperate or compete to stabilize excitonic order. The identification of broadly consistent experimental fingerprints enables clearer distinctions between excitonic order and structurally or correlation-driven alternatives. This unified framework provides a foundation for interpreting a rapidly expanding set of materials in which many-body electron–hole correlations play a central role. Looking ahead, the field is positioned for transformative developments. Dimensional engineering, moiré superlattices and topology-protected platforms offer routes to further enhance exciton binding energies and manipulate condensate properties with unprecedented precision, while ultrafast probes reveal pathways for coherent control on femtosecond timescales. Artificial bilayers and photoinduced condensates extend excitonic physics into regimes inaccessible in equilibrium, raising prospects for exciton-mediated superconductivity, tunable spin-coherent phases and device-relevant, low-dissipation functionalities. Continued progress in material synthesis, disorder control, and nanoscale engineering will be essential for realizing robust, electrically addressable excitonic architectures that harness macroscopic coherence as a resource for next-generation quantum and optoelectronic technologies.

# Display items

> **Box 1: Important models**
>
> The simplest lattice model that can exhibit excitonic order is the extended Falicov-Kimball model (EFKM) with the Hamiltonian:
>
> $$H_{\text{EFKM}} = \sum_{\langle i,j \rangle} -t_a a^\dagger_i a_j - t_b b^\dagger_i b_j + H.c.$$
> $$+ \sum_{\langle i,j \rangle} V_{ab} b^\dagger_i a_j + V_{ba} a^\dagger_i b_j + H.c. \quad (1)$$
> $$+ \frac{D}{2} \sum_i \left( a^\dagger_i a_i - b^\dagger_i b_i \right) + U' \sum_i a^\dagger_i a_i b^\dagger_i b_i$$
>
> where $a^\dagger_i$ and $b^\dagger_i$ are creation operators for spinless electrons that belong to two different orbitals or bands at the lattice site $\mathbf{R}_i$, $D$ is the energy difference between the two bands and $V_{ab}$ are the band hybridization terms. The difference between the extended and original Falicov-Kimball model is that the electrons in both orbitals are allowed to move with hopping integrals $t_a, t_b \neq 0$, which is crucial for the emergence of the excitonic order. Because of the spinless nature of the model, the only allowed on-site interaction is the interorbital term with strength characterized by $U'$.
>
> Inclusion of the spinful nature of electrons leads to the two-band Hubbard model with the Hamiltonian given by:
>
> $$H_{\text{2BHM}} = \sum_{\langle i,j \rangle,\sigma} -t_a a^\dagger_{i,\sigma} a_{j,\sigma} - t_b b^\dagger_{i,\sigma} b_{j,\sigma} + H.c.$$
> $$+ \sum_{\langle i,j \rangle,\sigma} V_{ab} b^\dagger_{i,\sigma} a_{j,\sigma} + V_{ba} a^\dagger_{i,\sigma} b_{j,\sigma} + H.c.$$
> $$+ \frac{D}{2} \sum_{i,\sigma} \left( a^\dagger_{i,\sigma} a_{i,\sigma} - b^\dagger_{i,\sigma} b_{i,\sigma} \right) + U' \sum_{i,\sigma,\sigma'} a^\dagger_{i,\sigma} a_{i,\sigma} b^\dagger_{i,\sigma'} b_{i,\sigma'} \quad (2)$$
> $$+ U \sum_i \left( a^\dagger_{i,\uparrow} a_{i,\downarrow} + b^\dagger_{i,\uparrow} b_{i,\downarrow} \right)$$
> $$- J \sum_{i,\sigma} \left( a^\dagger_{i,\sigma} a_{i,\sigma} b^\dagger_{i,\sigma} b_{i,\sigma} + a^\dagger_{i,\sigma} a_{i,\bar\sigma} b^\dagger_{i,\bar\sigma} b_{i,\sigma} \right)$$
> $$+ J' \sum_i a^\dagger_{i,\uparrow} a^\dagger_{i,\downarrow} b_{i,\downarrow} b_{i,\uparrow} + H.c.$$
>
> where $a^\dagger_{i,\sigma}$ and $b^\dagger_{i,\sigma}$ are corresponding creation operators of electrons with spin $\sigma$. Spinful electrons allow for more interaction terms, which now include intraorbital terms $U$, Hund's coupling $J$, and pair hopping $J'$. This model now allows for consideration of the spinful excitonic condensates, with the distinction between spin singlet and triplet order parameters.

**List of abbreviations**

| | |
|---|---|
| 2BHM | Two-Band Hubbard Model |
| 2D | Two-Dimensional |
| ARPES | Angle Resolved PhotoEmission Spectroscopy |
| BCS | Bardeen-Cooper-Schrieffer |
| BEC | Bose-Einstein Condensate |
| BZ | Brillouin Zone |
| CDW | Charge Density wave |
| DOS | Density of States |
| EFKM | Extended Falicov-Kimball Model |
| EI | Excitonic Insulator |
| FT | Fourier Transform |
| FFT | Fast Fourier Transform |
| LDOS | Local Density of States |
| PLD | Periodic Lattice Distortion |
| SDW | Spin Density Wave |
| STM | Scanning Tunnelling Microscopy |
| STS | Scanning Tunnelling Spectroscopy |
| SW | Spectral Weight |
| TMD | Transition Metal Dichalcogenide |
| VB | Valence Band |
| CB | Conduction Band |

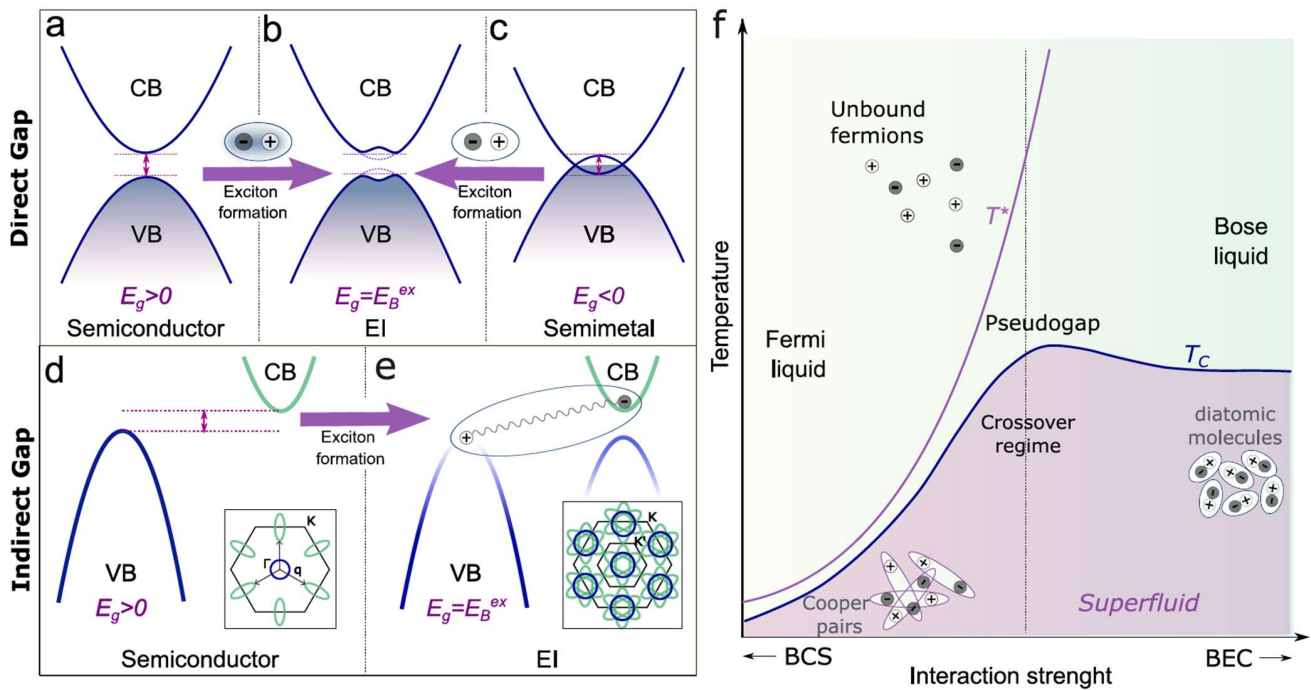

**Figure 1. The excitonic insulator concept.** Schematic representation of how electron–hole Coulomb attraction can drive excitonic condensation. Band structures are shown for a small gap semiconductor (a), a semimetal (c), and the resulting EI state with a hybridization gap (b), in the direct gap scenario. Schematic representation of the band structure and Fermi surface (inset) on an indirect gap semiconductor is depicted in panels (d), normal state, and (e), resulting excitonic state, showing SW redistribution, backfolded bands and reconstruction of the BZ. (f) Qualitative diagram illustrating the BCS–BEC crossover as a function of temperature and interaction strength, from weakly bound Cooper-like pairs (left) to tightly bound excitons (right). The pairing temperature $T*$, the and the onset of the crossover regime are also indicated.

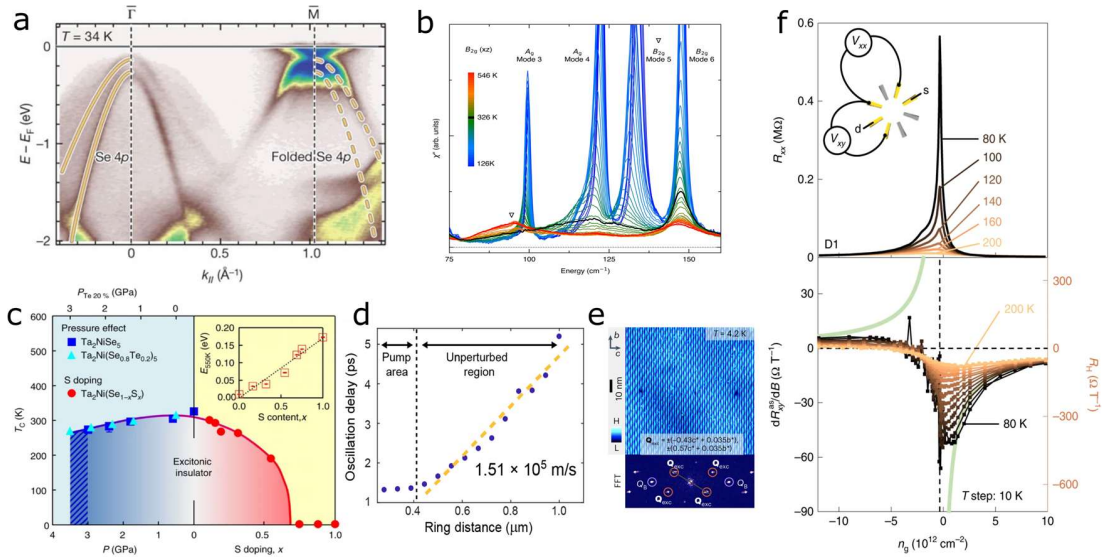

**Figure 2. Experimental signatures across multiple probes.** a| ARPES: Back-folded bands in 1T–TiSe$_2$; b| Raman spectroscopy: Phonon softening in Ta$_2$NiSe$_5$, where mode 5 (marked by an inverted triangle) exhibits a large blue shift upon cooling, and is heavily damped close to the $T_C$ (black solid line). c| Order melting: phase diagram of Ta$_2$NiSe$_5$ mapping the transition temperature $T_C$ as a function of S doping (right) and pressure and Te doping (left). d| Collective modes: Temporal characterization of the oscillatory modes' propagation in Ta$_2$NiSe$_5$ obtained by ultra-fast pump probe microscopy. e|Lattice effects: Appearance of a CDW below $T_C$ in Ta$_2$Pd$_3$Te$_5$. Top panel shows the topography obtained via STM (V=-200mV), the bottom panel its corresponding FFT, where well-defined wavevector peaks associated to the CDW (orange circles) are seen alongside the Bragg peaks (purple circles). f| Transport: Hall anomaly in monolayer WTe$_2$. Gate dependent $R_{xx}$ (top) and corresponding Hall resistance (bottom) temperatures: while the former diverges at lower temperatures, the latter remains finite. Panel **a** adapted from ref.[64], Springer Nature Limited. Panel **b** adapted from ref.[86], Springer Nature Limited. Panel **c** adapted from ref.[74], Springer Nature Limited. Panel **d** adapted from ref.[95], Science. Panel **e** adapted from ref.[165], Springer Nature Limited. Panel **f** adapted from ref.[20], Springer Nature Limited.

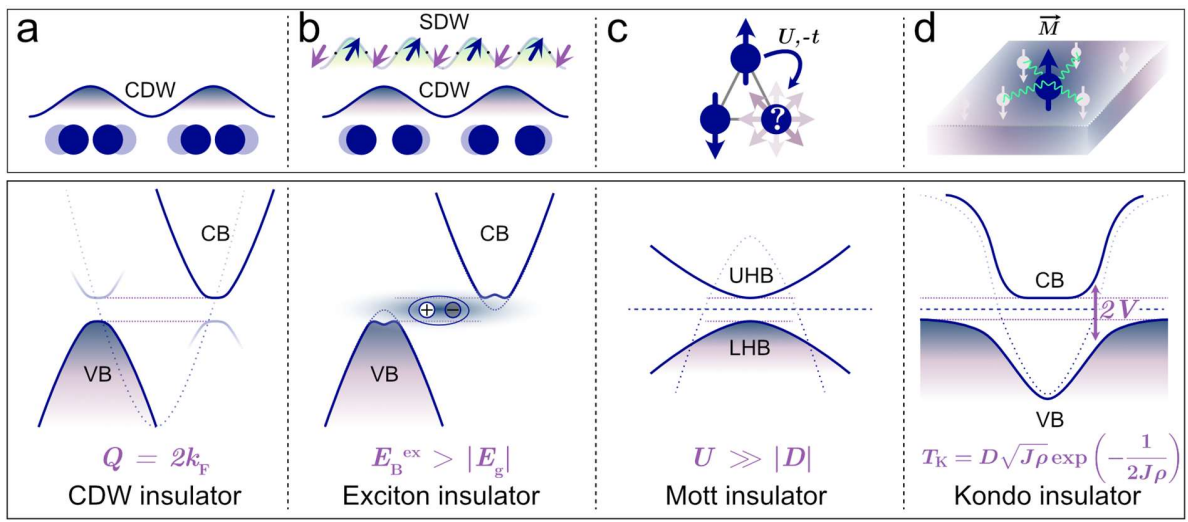

**Figure 3. Disentangling excitonic order from competing phases.** Comparative schematics of electronic order parameters to distinguish excitonic order from other correlated or symmetry-broken states. Schematic band dispersion and gap character for: (a) a charge density wave (gap opening tied to periodic lattice distortion and momentum-dependent folding); (b) an excitonic insulator (gap opening via exciton condensation, with minimal lattice distortion); (c) a Mott insulator (local Coulomb repulsion forms lower (LHB) and upper Hubbard bands (UHB) separated by gap, independent of Fermi surface nesting); (d) a Kondo/mixed-valence hybridization gap (f–d band mixing).

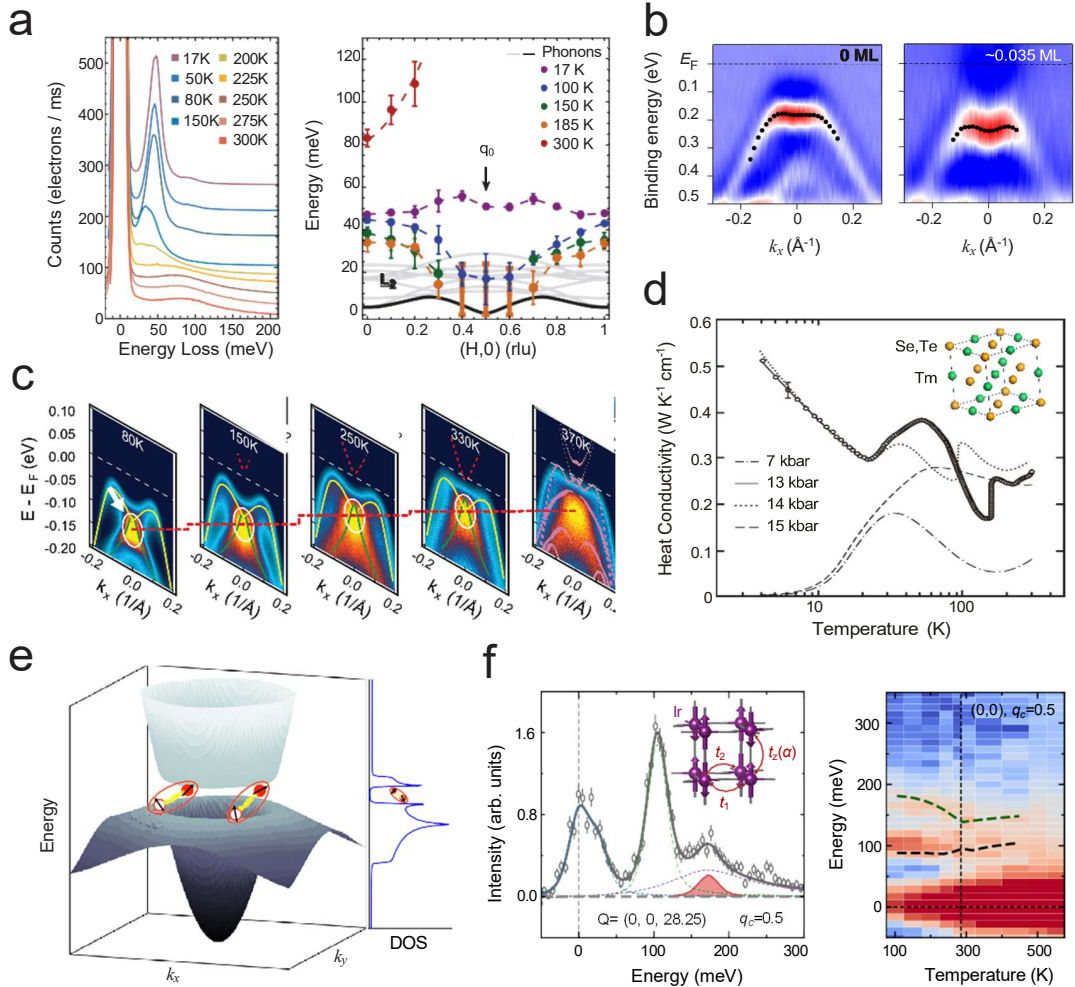

**Figure 4. Candidate materials: bulk canonical and correlated systems.** a| Softening of the plasmon mode in the excitonic transition of 1T-TiSe$_2$. Momentum-resolved electron energy loss spectroscopy (M-EELS) spectra at $q = 0$ reveal the temperature evolution of a collective mode that shifts to lower energy (softening) and sharpens upon cooling. Notably, such softening is mostly pronounced at $q_0 = 0.5$ across the transition temperature, evidencing the excitonic origin for the transition. b| Valence- band flattening in Ta$_2$NiTe$_5$. This band deformation can be tuned by potassium doping owing to the Stark effect near the surface, supporting its excitonic ground state. c| Temperature evolution of band gap in Ta$_2$Pd$_3$Te$_5$ revealed by ARPES. d| Heat conductivity anomalies in TmSe$_{0.45}$Te$_{0.55}$ under pressure, suggesting the possibility of a superfluid phase owing to the excitonic condensation. e| Excitons formation in topological Kondo insulator SmB$_6$. f| Excitonic mode condensation in antiferromagnetic insulator Sr$_3$Ir$_2$O$_7$. A longitudinal mode ( 170 meV) isolated in resonant inelastic x-ray scattering (RIXS) spectra exhibits distinct softening at the Néel temperature, evidencing the excitonic mode condensation. Panel a adapted from ref.[160], Science. Panel b adapted from ref.[122], APS. Panel c adapted from ref.[166], APS. Panel d adapted from ref.[185], APS. Panel e adapted from ref.[195], APS. Panel f adapted from ref.[200], Springer Nature Limited.

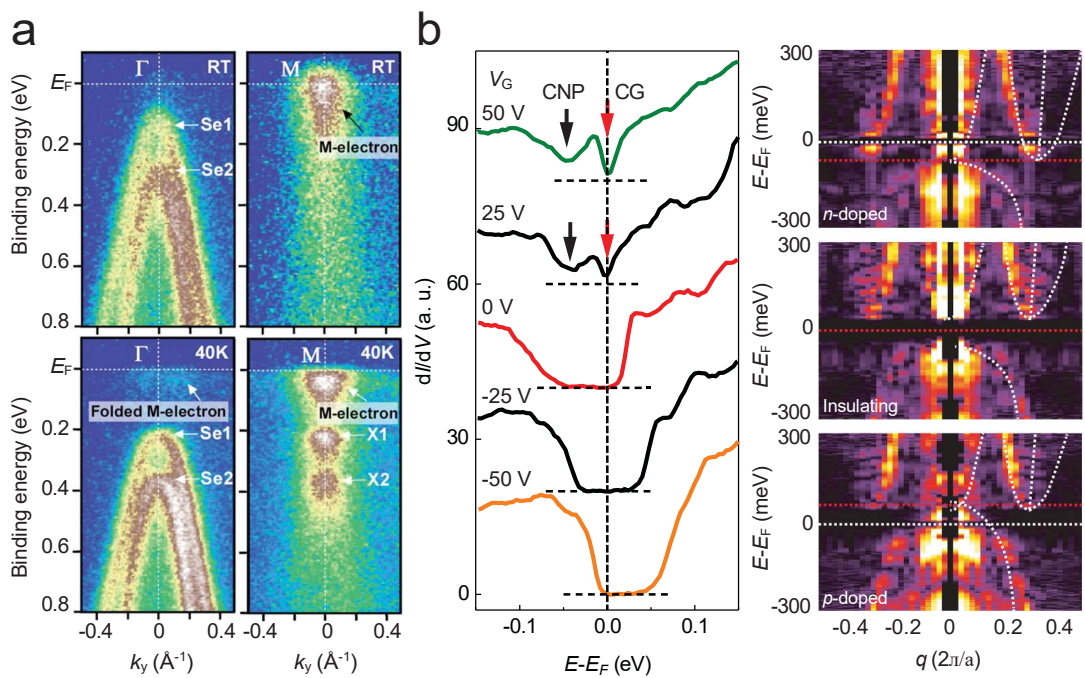

**Figure 5. Candidate materials: monolayer canonical and correlated systems.** a| Monolayer 1T-TiSe$_2$: Band folding and gap opening revealed by ARPES suggest an excitonic transition. b| Monolayer 1T'-WTe$_2$: A gate-tunable gap collapse observed by STS spectroscopy and quasiparticle interference measurements reveals an ambipolar quantum phase transition induced by electrostatic doping. Panel a adapted from ref.[66], ACS. Panel b: left adapted from ref.[79], Wiley, and right adapted from ref.[80], ACS.

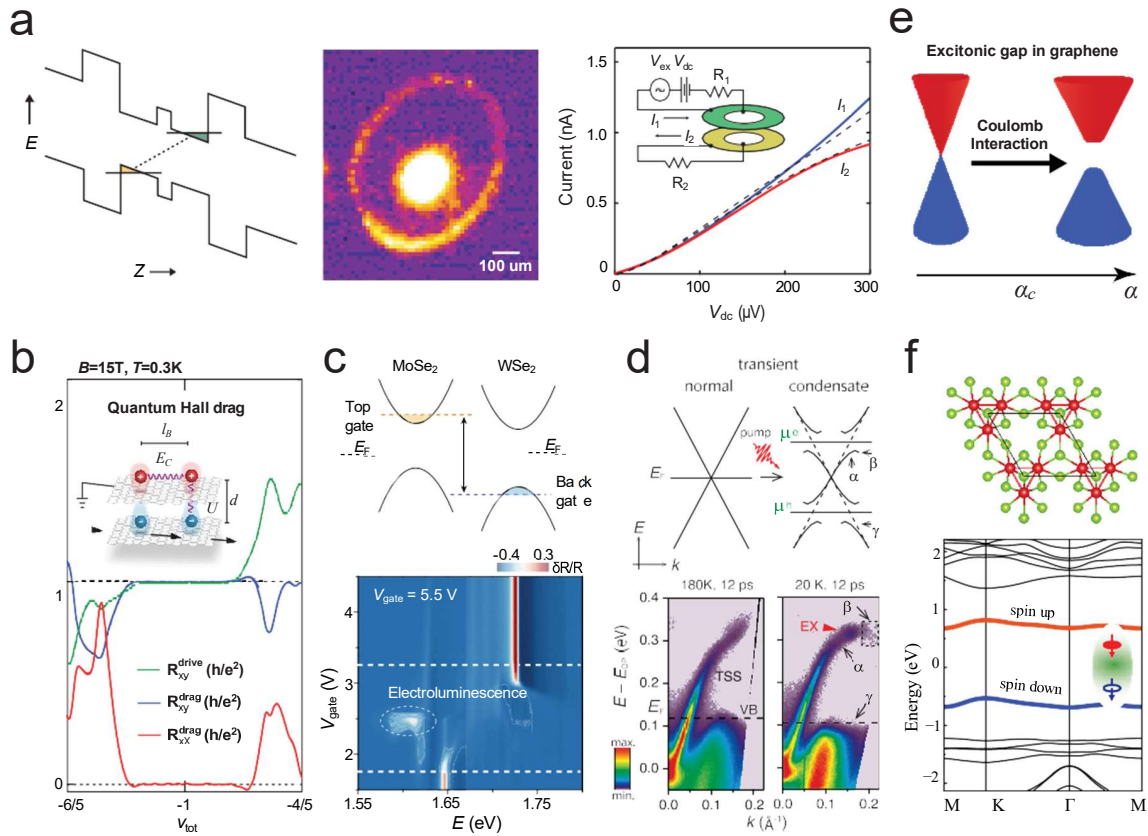

**Figure 6. Artificial, non-equilibrium and predicted excitonic platforms.** a| Semiconductor double quantum wells: An electron-hole bilayer system can be formed by applying a perpendicular electric field that tilted conduction and valence band edges (left panel), where Coulomb interactions can lead to exciton formation and condensate, as evidenced by the unique photoluminescence ring (middle panel) and perfect Coulomb drag effect (right panel). b| Electron-hole bilayer system in double graphene layers, where the superfluid states are evidenced by quantum Hall drag effect under strong magnetic fields. c| Electron-hole bilayer system in double TMD layers, where an extra electroluminescence is susceptible to doping, suggesting the condensate states of excitons. d| Transient excitonic condensation in Dirac bands: photo-excitations lead to excitonic gaps opening at the quasi-equilibrium chemical potentials and band edge renormalization (top panel), which are observed in time-resolved ARPES measurement (bottom panel). e| Excitonic transition in graphene: Strong Coulomb interaction gives rise to spontaneous mass generation through excitonic gap opening. f| Spin-triplet excitonic insulator in $Ta_3X_8$: The breathing Kagome lattice formed by Ta atoms (top panel) results in the formation of flat bands both in conduction and valence bands (bottom panel), which are spin polarized owing to strong SOC. Moreover, large binding energies gives rise to spontaneous formation and condensation of spin-triplet excitons. Panel a: left and middle panels adapted from ref.[202], Science; right panel adapted from ref.[207], Springer Nature Limited. Panel b adapted from ref.[211], Springer Nature Limited. Panel c adapted from ref.[213], Spinger Nature Limited. Panel d adapted from ref.[223], PNAS. Panel f adapted from ref.[24], licensed under CC BY 4.0.

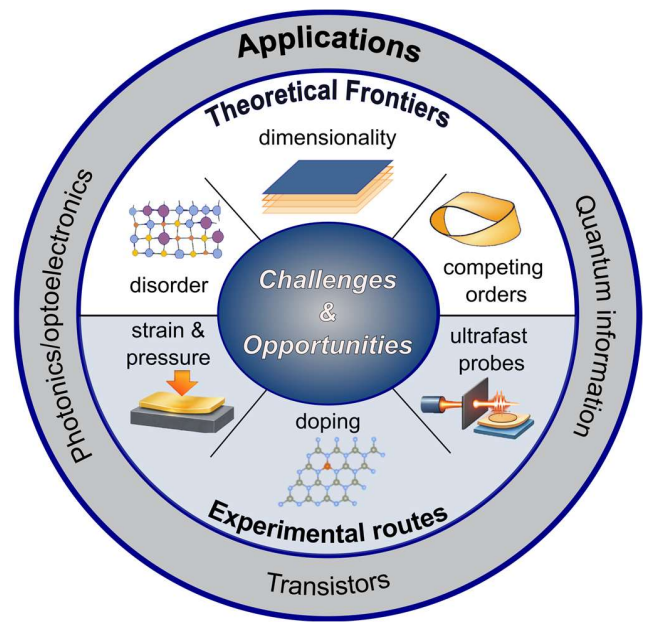

**Figure 7. Challenges and opportunities.** Schematic summarising key frontiers for the field.

**Table 1. Experimental fingerprints of excitonic insulators**

| Fingerprint | Description | Probes | Material examples |
|---|---|---|---|
| Gap opening | Temperature-dependent gap that opens below a critical temperature $T_c$ | ARPES, STS, Optical conductivity | TiSe$_2$[63, 65, 66], Ta$_2$NiSe$_5$, Ta$_2$Pd$_3$Te$_5$[18, 78], WTe$_2$[20, 79, 80], ZrTe$_2$[67, 68], HfTe$_2$[69] |
| Backfolding | Bandfolded band replicas at high symmetry points of the BZ, with SW redistribution | ARPES | 1T-TiSe$_2$[62–66], ZrTe$_2$[67, 68], HfTe$_2$[69] |
| VB flattening | Flattening of the VB top and shift towards higher binding energies | ARPES, STS | Ta$_2$NiSe$_5$[142], TiSe$_2$[63], ZrTe$_2$[67, 68], HfTe$_2$[69] |
| In-gap flat bands | Emergent flat sharp states inside the gap | ARPES, STS | Ta$_2$Pd$_3$Te$_5$[18, 78] |
| Persistence of excitonic features above $T_c$ | This indication suggests excitonic order fluctuations or preformed excitons | ARPES | TiSe$_2$[62,65,66], ZrTe$_2$[67], Ta$_2$NiSe$_5$[142] |
| Ultrafast electronic collapse | Fast loss of backfolded weight after pump (10–100 fs) | tr-ARPES, ultrafast optics | TiSe$_2$[64, 133], Ta$_2$NiSe$_5$[77] |
| Amplitude (Higgs) mode | Gapped collective excitation corresponding to fluctuations of the excitonic order parameter $\Delta$ | Raman, pump–probe, time-resolved spectroscopy | Ta$_2$NiSe$_5$[88–90], TiSe$_2$[92, 247] |
| Phase mode | Collective excitation associated with phase fluctuations $\varphi$ of the excitonic order parameter. | THz and IR spectroscopy, optical conductivity | Ta$_2$NiSe$_5$[53, 83] |
| CDW without PLD | Electronic CDW order in indirect-gap materials with no associated structural PLD | Raman, LEED, X-ray diffraction, neutron scattering, STM, TEM | ZrTe$_2$[68], HfTe$_2$[69], WTe$_2$[19,80] |
| Doping/gating gap collapse | Gap closes under chemical/electrostatic doping or gating, melting the excitonic order | ARPES, transport, gating, STS | WTe$_2$[79, 80], TiSe$_2$[118–120], Ta$_2$Pd$_3$Te$_5$[78], HfTe$_2$[69], Ta$_2$NiSe$_5$[121] |
| Pressure-induced CDW collapse | Phase-transition temperature is suppressed upon pressure application, melting the excitonic order | X-ray diffraction | TiSe$_2$[119,120,123], Ta$_2$NiSe$_5$[74, 124] |
| Drude and Hall carrier collapse | Drude weight loss as $T_c$ is approached and Hall carrier density drop below $T_c$ | Optical conductivity | TiSe$_2$[103,248], Ta$_2$NiSe$_5$[249] |
| Quantum oscillations | Oscillatory magnetotransport consistent with neutral, dipole-carrying, excitonic quasiparticles | Magnetotransport, capacitance and drag counterflow measurements | WTe$_2$[130], MoTe$_2$/WSe$_2$[131], MoSe$_2$/hBN/WSe$_2$[125,126] |

**Acknowledgements**
M.S.F. acknowledges support from the Australian Research Council Centre of Excellence FLEET (CE170100039). B.W. acknowledges financial support from the Singapore Ministry of Education (MOE) Academic Research Fund Tier 3 grant (MOE-MOET32023-0003) 'Quantum Geometric Advantage' and Singapore National Research Foundation (NRF) Grant (No. NRF-F-CRP-2024-0012). I.D.B. acknowledges support from the Ramon y Cajal program, grant no. YC2022-035562-I. IFIMAC acknowledges financial support from the Spanish Ministry of Science and Innovation through the "María de Maeztu" Programme for Units of Excellence in R&D (Grant CEX2023-001316-M). M.P. acknowledges financial support from The Robert A. Welch Foundation, Grant #L-E-0001-19921203. C.R. acknowledges support from the Ramon y Cajal program, grant no. YC2022-035562-I.

**Author contributions**
Y.Q. contributed to the writing of Sect. 3 and to figure creation.
C.R. wrote the Introduction, contributed to the writing of Sect. 2, created Table 1, and contributed to figure creation.
L.W. contributed to the writing of Sect. 2 and figure creation.
M.F. contributed to the discussion and to the editing of the manuscript in its final form.
M.P. contributed to the writing of Sect. 1.
B.W conceived the idea and outline together with IDB, contributed to the narrative development, and provided feedback and edits of the manuscript.
I.D.B. conceived the outline, coordinated the overall narrative development, integrated feedback across authors, contributed to figure creation and edited the manuscript.
All authors discussed the content, provided feedback during writing, and contributed to the final manuscript.

**Competing interests**
The authors declare no competing interests.

**Key Points**

- Excitonic insulators are many-body states where electron–hole pairing drives insulating behavior, distinct from band and Mott insulators, with coherent condensate properties.

- Identifying excitonic phases requires combining multiple experimental fingerprints due to competition and coexistence with other phases (e.g., charge density waves, Mott states).

- A broad range of materials families — from layered chalcogenides to engineered heterostructures — offer platforms to study and control excitonic phenomena.

- Future opportunities lie in ultrafast control, novel materials, and applications in low-dissipation electronics, quantum devices, and exciton-based technologies.